\documentclass[12pt]{article}
\usepackage{graphicx}
\usepackage{slashed}
\DeclareGraphicsExtensions{.pdf}
\usepackage{float} 
\usepackage{amsmath}
\usepackage{amsfonts}   
\usepackage{amssymb}

\begin{document}
\date{}
\title{{\bf{\Large Magnetoconductivity in chiral Lifshitz hydrodynamics}}}
\author{
 {\bf {\normalsize Dibakar Roychowdhury}$
$\thanks{E-mail:  dibakarphys@gmail.com}}\\
 {\normalsize  Indian Institute of Technology, Department of Physics,}\\
  {\normalsize Kanpur 208016, Uttar Pradesh, India}
}

\maketitle
\begin{abstract}
In this paper, based on the principles of linear response theory, we compute the longitudinal DC conductivity associated with Lifshitz like fixed points in the presence of chiral anomalies in ($ 3+1 $) dimensions. In our analysis, apart from having the usual anomalous contributions due to chiral anomaly, we observe an additional and pure \textit{parity odd} effect to the magnetoconductivity which has its origin in the broken Lorentz (boost) invariance at a Lifshitz fixed point. We also device a holographic set up in order to compute ($ z=2 $) Lifshitz contributions to the magnetoconductivity precisely at strong coupling and low charge density limit.
\end{abstract}

\section{Overview and Motivation}
\subsection{Chiral anomaly and Magnetoconductivity}
The various implications of quantum anomalies \cite{ch1}-\cite{ch3} on relativistic hydrodynamic systems \cite{ch4} has been an active area of research for the past couple of decades\footnote{As an example one might consider the hydrodynamic description of hot chiral QCD with two flavor d.o.f at temperatures much higher than the QCD transition temperature. In the absence of external magnetic field, both the isospin as well as the axial isospin currents are conserved. On the other hand, when an external magnetic field is switched on, the corresponding constitutive equations receive anomalous contributions and thereby the axial (isospin) current is no more conserved \cite{Newman:2005hd}. However, such a non conservation due to anomaly requires both the electric field as well as the magnetic field along the same direction. } \cite{Alekseev:1998ds}-\cite{Newman:2005hd}. The fact that is well understood by now is that the presence of anomalies always induce transport processes in a relativistic hydrodynamic system without any effects of dissipation and thereby they do not contribute to the \textit{local} entropy production with in the system itself \cite{Landsteiner:2012kd}-\cite{Banerjee:2014cya}. Keeping the spirit of this intriguing fact, it is noteworthy to mention that the physics of (chiral) anomalies has attained renewed attention for the past one decade in the context of relativistic heavy ion collisions where during the early non equilibrium stages of the collision one might have an imbalance between the number of left handed and right handed quarks \cite{Fukushima:2008xe}-\cite{Kharzeev:2007jp}. This phenomena leads to so called axial anomalies which trigger an electric current in the presence of an external magnetic field \cite{Fukushima:2008xe}-\cite{Kharzeev:2007jp}. 

Chiral anomalies in $ (3+1) $ dimensions lead to anomalous transports of two types at the level of the first order dissipative hydrodynamics. The first one is known as the chiral magnetic conductivity, which is the transport associated with electrical conductivity parallel to the direction of the magnetic filed. The other (anomalous) transport associated with chiral anomalies is known as the chiral vortical conductivity \cite{Erdmenger:2008rm}-\cite{Banerjee:2008th} which is the transport associated with the induced current sourced due to the vortices present in the fluid \cite{Erdmenger:2008rm}-\cite{Banerjee:2008th}. Keeping these facts in mind, one could in principle express the constitutive relations corresponding to a $ U(1) $ charged anomalous fluid in the Landau frame as\footnote{At this stage it is noteworthy to mention that considering a most general approach, one could in principle construct anomalous hydrodynamics with $ n $ number of anomalous charges ($ \varrho^{(n)} $). However, physically the most interesting situation arises when we consider the case with $ n=2 $ $ U(1) $ charges. In that case one needs to define an axial vector current ($ J^{\mu}_{5} $) and the vector current ($ J^{\mu} $) where the later one is guaranteed to be conserved by means of the Bardeen counter term. For the purpose of our present computations, we would however stick to the $ n=1 $ case where the $ U(1) $ current that we consider plays the analogous role of the axial vector current \cite{Gahramanov:2012wz}. } \cite{Son:2009tf},
\begin{eqnarray}
T^{\mu\nu}&=& (\epsilon + p)u^{\mu}u^{\nu}+pg^{\mu \nu}-\eta P^{\mu \alpha}P^{\nu\beta}(\nabla_{\alpha}u_{\beta}+\nabla_{\beta}u_{\alpha})-\left(\zeta -\frac{2}{3}\eta \right) P^{\mu \nu}(\nabla . u)\nonumber\\
J^{\mu}&=&\varrho u^{\mu} + \sigma_{E}\left( E^{\mu}-T P^{\mu \nu}\nabla_{\nu}\left(\frac{\mu}{T} \right) \right) +\sigma_{B}B^{\mu}+\sigma_{V}\omega^{\mu}
\end{eqnarray}
where, $ P^{\mu \nu}=g^{\mu \nu}+u^{\mu}u^{\nu} $ is the usual projection operator and $ \sigma_{B} $ and $ \sigma_{V} $ are respectively the chiral magnetic and chiral vortical conductivities.

In \cite{Son:2009tf}, the authors had shown that these two transports could in principle be fixed almost uniquely within the theory itself by demanding the positive definiteness of the \textit{local} entropy current. However, it was shown later on that depending on the temperature of the system one could still add various other terms to these anomalous transport coefficients as undetermined integration constants \cite{Neiman:2010zi}. Keeping the spirit of our current discussion, it is noteworthy to mention that the computation of the anomalous transports directly by using the Kubo formula has been initiated recently \cite{Amado:2011zx}-\cite{Gynther:2010ed} where the analysis has been extended in order to incorporate the effects of mixed gauge-gravitational anomalies in four dimensions \cite{Landsteiner:2011cp}-\cite{Megias:2013xla}. At this stage it is noteworthy to mention that the Kubo formulae for anomalous transports are in fact quite different from that of the usual Kubo formulae for dissipative transports. In order to evaluate anomalous transports using Kubo formulae, one first needs to take the zero frequency limit and then the zero momentum limit. Whereas on the other hand, in case of dissipative transports it works the other way around \cite{Amado:2011zx}.

The effects of mixed gauge-gravitational anomalies in four dimensions appear in a strange manner. The reason for this rests on the fact that although these effects are higher order in derivatives, still they show up as a purely temperature dependent effect at the first order level in the derivative expansion. However, at this stage it is customary to note the following fact: In case of pure chiral anomalies, using the notion of positive definiteness of the local entropy production one could in principle fix the corresponding transports associated with it. On the other hand, no such analogous method has been developed yet in order to fix the transports associated with mixed gauge-gravitational anomalies.

One of the intriguing features of chiral anomalies in ($ 3+1 $) dimensions is the existence of the longitudinal magnetoconductivity along the direction of the external magnetic field \cite{Nielsen:1983rb}-\cite{Jimenez-Alba:2015awa}. In other words, in the presence of chiral anomalies, there appear to be additional contributions to the longitudinal DC electrical conductivity along the direction of the background magnetic field. This additional contribution strongly enhances the value of the DC electrical conductivity from its usual value. The existence of negative magnetoresistivity in Weyl metals has been investigated extensively in \cite{Son:2012bg}. This analysis eventually hints towards an experimental realization of axial anomalies under a solid state set up. In their analysis \cite{Son:2012bg}, the authors claim that although the negative magnetoresistivity is concerned with the triangle anomaly, still it takes place at the classical level where one could ignore the electron mean free path compared to the magnetic length scale of the theory. In the following we summarize various characteristic features of longitudinal magnetoconductivity \cite{Landsteiner:2014vua} :\\ 
$ \bullet $ In the context of relativistic ($ z=1 $) hydrodynamics, this effect is solely generated by the chiral anomaly itself, which therefore disappears in the absence of the anomaly.\\ 
$ \bullet $ The contributions coming from various dissipative effects (for example, the energy relaxation, the charge relaxation and the momentum relaxation) present in the system play crucial role in order to generate a finite longitudinal DC magnetoconductivity within the system.\\ 
$ \bullet $ Moreover, in \cite{Landsteiner:2014vua}, considering the zero charge density limit, the authors had computed the anomalous contribution to the DC electrical conductivity under a holographic set up. These computations eventually correspond to longitudinal magnetoconductivity in a strongly coupled system. From their analysis, one could easily notice that the anomalous contribution to the conductivity goes as, $ \sim T^{-2} $ and is proportional to the square of the external magnetic field.

\subsection{Hydrodynamics at a Lifshitz fixed point}
For the past few years, the hydrodynamic description of quantum critical systems with Lifshitz scaling symmetry \cite{Hoyos:2013eza}-\cite{Hoyos:2015lra} has been an active area of research due to its several remarkable features among which the most significant one is the description of strange metals near the quantum criticality where the usual Landau Fermi liquid theory does not hold good. Quantum critical points are believed to be the best candidates in order to describe several physical properties of heavy Fermion compounds including the high $ T_{c} $ superconductors. The hydrodynamic description of Lifshitz like fixed points exists under certain limiting conditions namely, when the length scale ($ l_{T}\sim T^{-1/z} $) associated with thermal fluctuations is quite small compared to that of the correlation length ($ \xi \gg l_{T} $) of the theory. This hydrodynamic sector also covers part of the superconducting dome where the symmetry is broken spontaneously.

Lifshitz fixed points are always special in the sense that Lifshitz symmetry algebra does not include the generators of the Lorentz boost symmetry which eventually results in a number of additional transports in the hydrodynamic description of the theory. This point could be further elaborated as follows. We know that the divergence of the Noether current ($ \mathcal{M}^{\mu\nu\lambda} $) associated with Lorentz invariance could be formally expressed as,
\begin{eqnarray}
\partial_{\mu}\mathcal{M}^{\mu\nu\lambda}=T^{\nu\lambda}-T^{\lambda\nu}.
\label{eqn2}
\end{eqnarray}
It is therefore quite evident from the above equation (\ref{eqn2}), that if the corresponding current associated with Lorentz transformations is not conserved then the stress tensor need not necessarily have to be symmetric namely, $T^{\nu\lambda} \neq T^{\lambda\nu} $.

The role of parity breaking transports (in ($ 3+1 $) dimensions) on the hydrodynamic sector of the Lifshitz like fixed points has been explored very recently in \cite{Hoyos:2015lra}. There the authors study the effect of chiral anomalies on the hydrodynamic transports in the presence of the broken Lorentz boost invariance. From the experimental point of view such a theoretical attempt is important due to the fact that most of the strange metal phenomena are observed either in magnetic materials or in the presence of the external magnetic field. In there analysis, the authors have found that apart from having the usual parity odd transports namely, the chiral magnetic conductivity as well the chiral vortical conductivity one encounters additional (non)dissipative transports (due to the broken Lorentz symmetry) which could be uniquely fixed by demanding the positivity of the local entropy production \cite{Hoyos:2015lra}.

\subsection{Our goal}
Keeping the spirit of the discussions made so far, the purpose of the present article is to carry out an explicit analytic computation of the magnetoconductivity associated with Lifshitz like fixed points under the framework of the so called \textit{linear response} theory. From the experimental point of view, some of the crucial theoretical predictions of our analysis should be testable in various solid state set up (particularly in strange metal systems\footnote{As for example, one could list certain ferromagnetic materials like, $ MnSi $, $ ZrZn_2 $, unconventional cuprate superconductors and iron pnictides, electronic nematics like $ Sr_3Ru_2O_7 $ etc. \cite{sm1}-\cite{sm2}.}) in the near future. In our analysis, we are particularly interested to explore the following issues.\\
$ \bullet $ The primary concern of our analysis would be to explore the Lifshitz sector (we call it as $  \Theta_{\mathfrak{L}} $) of the longitudinal DC electrical conductivity ($ \sigma_{DC} $) and in particular how the effects of chiral anomalies enter into this sector.\\
$ \bullet $ The second motivation of our analysis would be to study the effects of relaxation time in this sector and in particular the behavior of $ \Theta_{\mathfrak{L}} $ in the low frequency ($ \mathfrak{w} $) limit i.e, whether there exists any pole in the limit $ \mathfrak{w} \rightarrow 0$.\\
$ \bullet $ Finally, our aim would be to sketch a holographic set up in order to evaluate the entity $ \Theta_{\mathfrak{L}} $ under certain specific assumptions namely, in the limit of the zero charge density ($ \varrho $) and at high temperatures ($ T $). This formulation would eventually provide us with some basic characteristic features (like the scaling of $ \Theta_{\mathfrak{L}} $ with temperature) of $ \Theta_{\mathfrak{L}} $ at strong coupling.

The organization of the paper is the following. In Section 2, we review the parity odd Lifshitz hydrodynamics in ($ 3+1 $) dimensions. In Section 3, we compute the magnetoconductivity for Lifshitz like fixed points in ($ 3+1 $) dimensions. In Section 4, we provide a holographic platform in order to evaluate the Lifshitz contribution ($ \Theta_{\mathfrak{L}} $) at strong coupling. Finally, we conclude in Section 5.

\section{Lifshitz hydrodynamics}
\subsection{Uncharged fluid}
We start our analysis with the formal introduction to the basic characteristic features of Lifshitz hydrodynamics in general in $ d+1 $ dimensions \cite{Hoyos:2013eza}. Hydrodynamic systems with Lifshitz scale invariance differ significantly from that of the usual (Lorent invariant) relativistic hydrodynamic systems due to the presence of the new transport coefficients allowed by the lack of (Lorentz) boost invariance. Due to the presence of the rotational invariance, these additional transports could be attributed starting from the first order dissipative level in the constitutive relation of the stress tensor namely \cite{Hoyos:2013eza}, 
\begin{eqnarray}
T^{\mu \nu}=\epsilon u^{\mu}u^{\nu}+p P^{\mu \nu}+\Pi^{(\mu \nu)}_{S}+\Pi^{[\mu \nu]}_{A}+(u^{\mu}\Pi^{[ \nu \sigma]}_{A}+u^{\nu}\Pi^{[\mu \sigma]}_{A})u_{\sigma}\label{E1}
\end{eqnarray}
where, $ \epsilon $ and $ p $ are respectively the energy and the pressure density and $ P^{\mu \nu}=\eta^{\mu \nu}+u^{\mu}u^{\nu} $ is the so called projection operator\footnote{Here $ u^{\mu} $ is the four velocity such that $ u^{\mu}u_{\mu}=-1 $.}. Note that here $ \Pi^{(\mu \nu)}_{S} $ and $ \Pi^{[\mu \nu]}_{A} $ are respectively the symmetric as well as the antisymmetric combination of the dissipative terms at the level of the first order derivative expansion whose details will be fixed going into certain specific frame of reference as well as by imposing constraints due to the second law of thermodynamics.

In our analysis, we restrict ourselves to Landau frames namely, $ T^{\mu \nu}u_{\nu}=-\epsilon u^{\mu} $. This eventually constrains the form of dissipative terms in the constitutive relation (\ref{E1}). For example, the symmetric part of the dissipation must satisfy the condition $ \Pi^{(\mu \nu)}_{S}u_{\nu}=0 $. On the other hand, the anti symmetric part takes the form, $ \Pi^{[\mu \nu]}_{A}=u^{[\mu}V^{\nu]}_{A} =\frac{1}{2}\left(u^{\mu}V^{\nu}_{A}- u^{\nu}V^{\mu}_{A}\right) $ such that $ V^{\nu}_{A} u_{\nu}=0$. Keeping these facts in mind, the energy momentum tensor of an uncharged (Lifshitz) fluid in the Landau frame takes the following form \cite{Hoyos:2015lra},
\begin{eqnarray}
T^{\mu \nu}= \epsilon u^{\mu}u^{\nu}+p P^{\mu \nu}+\Pi^{(\mu \nu)}_{S}+u^{\mu}V^{\nu}_{A}.
\end{eqnarray}

\subsection{Charged fluid}
In order to define charged fluids with Lifshitz scaling symmetry, one needs to consider an additional constitutive relation for the $ U(1) $ charged current namely \cite{Hoyos:2015lra},
\begin{eqnarray}
J^{\mu}=\varrho u^{\mu}+\Gamma^{\mu}
\end{eqnarray}
where $ \Gamma^{\mu} $ is the full first order dissipative correction to the charge current that contains all the terms at the level of the first order in the derivative expansion\footnote{In principle $ \Gamma^{\mu} $ contains the full set of dissipative corrections due to both parity even as well as parity odd terms in the constitutive relation at the level of the first order derivative expansion. However in this section we only consider contributions coming from the parity even sector which we denote as $ \Gamma^{\mu}_{P} $. The parity odd sector will be included in the next section.}. Following our previous arguments, in the Landau frame we are supposed to impose certain constraints over the dissipative corrections to the charge current, namely $ \Gamma^{\mu}u_{\mu}=0 $, which thereby determines the charge density as, $ \varrho =-J^{\mu}u_{\mu} $.

Before we proceed further, it is important to have some discussions on the scaling dimensions of various thermodynamic entities and/or physical parameters in a theory with Lifshitz scaling symmetry. For a theory with the Lifshitz scale invariance,
\begin{eqnarray}
t \rightarrow \lambda^{z} t,~~x^{i} \rightarrow \lambda x^{i},~~i=1,....,d
\end{eqnarray}
the temperature ($ T $) as well the chemical potential ($ \mu $) scales as, $ [T]=[\mu]=z $. The speed of light on the other hand is considered to be a quantity with non zero scaling dimensions namely\footnote{In particular, for field theories at a Lifshitz fixed point ($ z\geq 2 $), the Lorentz invariance is explicitly broken near the UV scale of the theory and as a result the speed of light at that scale may turn up to infinity. This leads to the modification of the so called dispersion relation at short distances namely, $ v_{g}\sim z\left(\frac{\mathfrak{q}}{M} \right)^{z-1}  $, where, $ \mathfrak{q} $ is the spatial momentum and $ M $ is the mass of the scalar particle such that, $ \mathfrak{q}>>M $ near the UV scale of the theory \cite{Chen:2009ka}. For such theories, however, the Lorentz invariance could be restored back at large distances where one could (approximately) set the speed of light equal to unity.}, $ [c]=z-1 $. The scaling dimensions corresponding to the rest of the parameters of the theory turn out to be,
\begin{eqnarray}
[u^{\mu}]=0,~~[\epsilon]=[p]=z+d,~~[\varrho]=d.
\end{eqnarray}

Like in the uncharged case, in the constitutive relations corresponding to a charged viscous fluid one can in principle add all possible terms that come up with derivatives of the fluid velocity ($ u^{\mu} $), temperature ($ T $), chemical potential ($ \mu $) and the gauge field ($ A_{\mu} $). Not all of these terms are physically relevant and they turn out to be highly constrained due to the second law of thermodynamics which could be expressed mathematically in its local form as,
\begin{eqnarray}
\partial_{\mu}j^{\mu}_{s} \geq 0
\end{eqnarray}
where,  
\begin{eqnarray}
j^{\mu}_{s} = s u^{\mu} -  \frac{\mu}{T}\Gamma^{\mu}
\end{eqnarray}
is the so called entropy current where $ s $ is the canonical entropy density that obeys the Euler relation,
\begin{eqnarray}
\epsilon + p = T s + \mu \varrho \label{E8}
\end{eqnarray}
as well as the first law,
\begin{eqnarray}
\delta \epsilon = T \delta s + \mu \delta \varrho.
\label{E9}
\end{eqnarray}
After some straightforward calculations, the most general (parity even) viscous terms allowed by the second law of thermodynamics could be formally expressed as \cite{Hoyos:2015lra},
\begin{eqnarray}
T^{\mu \nu}= \epsilon u^{\mu}u^{\nu}+p P^{\mu \nu} - \eta P^{\mu \alpha}P^{\nu \beta}\left(\partial_{\alpha}u_{\beta}+ \partial_{\beta}u_{\alpha}-\frac{2}{d}P_{\alpha \beta}(\partial. u)\right) - \zeta P^{\mu \nu}(\partial. u)\nonumber\\
- \alpha_{1}u^{\mu}a^{\nu} - 2 \alpha_{2}u^{\mu}E^{\nu}+2 \alpha_{2}T u^{\mu}P^{\nu \sigma}\partial_{\sigma}\left( \frac{\mu}{T}\right)
\end{eqnarray}
\begin{eqnarray}
J^{\mu}=\varrho u^{\mu} + 2\alpha_{3}a^{\mu} +\sigma_{E} E^{\mu}-\sigma_{E}T P^{\mu \sigma}\partial_{\sigma}\left(\frac{\mu}{T} \right) 
\end{eqnarray}
where, $ E^{\mu}=F^{\mu \nu}u_{\nu} $ is the external electric field strength and $ a^{\mu}=u^{\sigma}\partial_{\sigma}u^{\mu} $ is the acceleration. Finally, it is the positivity of the entropy current that essentially imposes the following constraints on the transport coefficients namely \cite{Hoyos:2015lra},
\begin{eqnarray}
\eta \geq 0,~~ \zeta \geq 0, ~~\sigma_{E} \geq 0, ~~\sigma_{E}\alpha_{1} \geq (\alpha_{2}+\alpha_{3})^{2}.
\end{eqnarray}

\subsection{Parity odd transport}
We now restrict ourselves to $ 3+1 $ dimensions and extend the above formalism by considering the parity odd transports that one could possibly add (at the level of the first order derivative expansion) to the constitutive relations of the stress tensor ($ T^{\mu \nu} $) as well as the charge current ($ J^{\mu} $). These parity odd transports have their origin into the chiral anomalies associated with Weyl fermions in $ (3+1) $ dimensions namely \cite{Son:2009tf},
\begin{eqnarray}
\partial_{\mu}J^{\mu}=\mathfrak{c} E_{\mu}B^{\mu}.
\end{eqnarray}

The resulting parity odd terms that one might add to the stress tensor ($ T^{\mu \nu} $) as well as the charge current ($ J^{\mu} $) could be formally expressed as \cite{Hoyos:2015lra},
\begin{eqnarray}
T^{\mu \nu}_{\slashed P}&=&-T \beta_{\omega}u^{\mu}\omega^{\nu}-T\beta_{B}u^{\mu}B^{\nu}\nonumber\\
\Gamma^{\mu}_{\slashed P}&=&\sigma_{V}\omega^{\mu}+\sigma_{B}B^{\mu}.
\end{eqnarray}
Note that here $ \omega^{\mu} $ and $ B^{\mu} $ are respectively the vorticity and the external magnetic field strength that could be formally expressed as \cite{Son:2009tf},
\begin{eqnarray}
\omega^{\mu}=\frac{1}{2}\varepsilon^{\mu \nu \rho \sigma}u_{\nu}\partial_{\rho}u_{\sigma},~~~B^{\mu}= \frac{1}{2}\varepsilon^{\mu \nu \rho \sigma}u_{\nu}F_{\rho \sigma}.
\end{eqnarray}

The corresponding entropy current could be readily expressed as,
\begin{eqnarray}
j^{\mu}_{s}=s u^{\mu}-\frac{\mu}{T}\Gamma^{\mu}+D_{V}\omega^{\mu}+D_{B}B^{\mu}\label{E16}
\end{eqnarray}
where the last two terms on the r.h.s. of (\ref{E16}) are imposed by hand in order to ensure the positivity of the entropy current \cite{Son:2009tf}. From (\ref{E16}), it is in fact quite trivial to compute the divergence of the local entropy current which for the present case turns out to be \cite{Hoyos:2015lra},
\begin{eqnarray}
\partial_{\mu}j^{\mu}_{s} = \Delta_{P}+\Delta_{\slashed P}
\label{e17}
\end{eqnarray}
where,
\begin{eqnarray}
\Delta_{P} = -\frac{1}{T}\pi_{S}^{(\mu \nu)}\partial_{\mu}u_{\nu}-\frac{1}{T}V^{\mu}_{A}a_{\mu}+\frac{1}{T}\Gamma^{\mu}_{P}\left(E_{\mu}-T P_{\mu}\ ^{\nu}\partial_{\nu}\left( \frac{\mu}{T}\right)\right) 
\end{eqnarray}
is the contribution of the parity even sector to the entropy current and,
\begin{eqnarray}
\Delta_{\slashed P} = -\frac{1}{T}V^{\mu }_{A,\slashed P}a_{\mu}+\frac{1}{T}\Gamma^{\mu}_{\slashed P}\left(E_{\mu}-T P_{\mu}\ ^{\nu}\partial_{\nu}\left( \frac{\mu}{T}\right)  \right)-\frac{\mathfrak{c}\mu}{T}E_{\mu}B^{\mu}+\partial_{\mu}(D_{V}\omega^{\mu})+\partial_{\mu}(D_{B}B^{\mu}) 
\end{eqnarray}
is the contribution that comes from the parity odd sector\footnote{For details see \cite{Hoyos:2015lra}.}.

As reported by the authors in \cite{Hoyos:2015lra}, the positivity of the entropy current could be ensured iff the parity odd terms cancel among themselves. This cancellation could be realized \textit{on-shell} i.e; by using the ideal hydrodynamic equations. This also fixes the acceleration as \cite{Hoyos:2015lra},
\begin{eqnarray}
a^{\mu}=\frac{\varrho}{\epsilon +p}E^{\mu}-\frac{1}{\epsilon +p}P^{\mu \nu}\partial_{\nu}p.\label{E17}
\end{eqnarray} 

Finally, by computing $ \partial_{\mu}j^{\mu}_{s} $ from (\ref{e17}) and demanding the positivity of the entropy current one essentially ends up with the following set of solutions corresponding to the parity odd transports in $ 3+1 $ dimensions in the presence of the Lifshitz scale invariance\footnote{The enthusiastic reader should consult \cite{Hoyos:2015lra} for the details of the derivation.} \cite{Hoyos:2015lra},
\begin{eqnarray}
\beta_{\omega}&=&4\gamma_{B}\mu T,~~\beta_{B}=0\nonumber\\
\sigma_{B}&=&\mathfrak{c} \left( \mu - \frac{\varrho \mu^{2}}{2(\epsilon + p)}\right) -\frac{\varrho T^{2}}{\epsilon +p}\gamma_{B}\nonumber\\
\sigma_{V}&=&\mathfrak{c} \left( \mu^{2} - \frac{2\varrho \mu^{3}}{3(\epsilon + p)}\right)+2\gamma_{B}T^{2}\left( 1 -\frac{\mu \varrho}{\epsilon +p}\right) -2\gamma_{\omega}\frac{\varrho T^{3}}{\epsilon +p}.
\end{eqnarray}

Combining all these pieces of information together, the complete set of constitutive relations (upto the first order in the derivative expansion) for a charged anomalous fluid at its Lifshitz fixed point could be formally expressed as\footnote{By demanding the CPT invariance we would finally set, $ \gamma_{\omega}=0 $ \cite{Hoyos:2015lra}.},
\begin{eqnarray}
T^{\mu \nu}= \epsilon u^{\mu}u^{\nu}+p P^{\mu \nu} - \eta P^{\mu \alpha}P^{\nu \beta}\left(\partial_{\alpha}u_{\beta}+ \partial_{\beta}u_{\alpha}-\frac{2}{3}P_{\alpha \beta}(\partial. u)\right) - \zeta P^{\mu \nu}(\partial. u)\nonumber\\
- \alpha_{1}u^{\mu}a^{\nu} - 2 \alpha_{2}u^{\mu}E^{\nu}+2 \alpha_{2}T u^{\mu}P^{\nu \sigma}\partial_{\sigma}\left( \frac{\mu}{T}\right)-4\gamma_{B}\mu T^{2}u^{\mu}\omega^{\nu}\label{E19}
\end{eqnarray}
\begin{eqnarray}
J^{\mu}=\varrho u^{\mu} + 2\alpha_{3}a^{\mu} +\sigma_{E} E^{\mu}-\sigma_{E}T P^{\mu \sigma}\partial_{\sigma}\left(\frac{\mu}{T} \right)+ \sigma_{V}\omega^{\mu}+\sigma_{B}B^{\mu}.\label{E20}
\end{eqnarray}
Eqs (\ref{E19}) and (\ref{E20}) together with (\ref{E17}) is precisely the starting point of our analysis.

\section{Conductivity at Lifshitz fixed point }
With the above set up in hand, the purpose of the present section is to carry out an explicit computation for the anomalous longitudinal DC conductivity \cite{Landsteiner:2014vua} along the direction of the external magnetic field in the presence of the Lifshitz scaling symmetry. We start our analysis with the following assumptions that in the hydrodynamic limit one can generally claim $ T \geq \mu $, $ E \ll T^{2} $ and $ |\mathfrak{c}B|\ll T^{2} $ so that it is sufficient to consider terms upto leading order in the derivative expansion and ignore all the higher order terms. 

To start with we consider our system to be in equilibrium in the so called grand canonical ensemble  which is characterized by the Euler relation (\ref{E8}) together with the following identity,
\begin{eqnarray}
dp = s dT+\varrho d \mu \label{E21}
\end{eqnarray}
whose density matrix could be characterized in terms of two physical parameters namely, the temperature ($ T $) and the chemical potential ($ \mu $) along with a specific choice for the velocity vector field namely, $ u^{\mu}=(1,0,0,0) $. Moreover, we assume that initially there is a background magnetic field ($ \mathfrak{B} $) along the $ z $ direction and there is no background electric field ($ E^{\mu}=F^{t \mu}=0 $) to start with. Under such circumstances different components of the stress tensor as well as the $ U(1) $ current could be formally expressed as,
\begin{eqnarray}
T^{tt}=\epsilon (\mu , T), ~T^{ti}=0,~T^{ii}=p(\mu , T),~J^{t}=\varrho(\mu , T),~ J^{z}=\sigma_{B} \mathfrak{B}.
\end{eqnarray}

We now perturb our system by turning on an external electric field $ \delta E^{\mu} $ and we focus particularly on the longitudinal component of the electric field ($ \delta E_{z} $) since this is the only component that is responsible for the anomalous contribution in the DC conductivity. In the following we enumerate all possible perturbations in the system upto linear order in the fluctuations,
\begin{eqnarray}
\delta E^{z}&=& \delta F^{tz},~\delta E^{x}=\delta F^{tx}+ \mathfrak{B} \delta u_{y},~\delta E^{y}=\delta F^{ty}- \mathfrak{B}\delta u_{x}\nonumber\\
\mu (t, \textbf{x})&=& \mu + \delta \mu (t,\textbf{x}),~ T(t,\textbf{x})= T+\delta T(t, \textbf{x}),~ u^{\mu}(t, \textbf{x})=(1, \delta u_{i}(t,\textbf{x})).\label{E23}
\end{eqnarray}

With the above perturbations (\ref{E23}) in hand, it is in fact quite straightforward to show that the following relations are true upto leading order in the fluctuations,
\begin{eqnarray}
\delta T^{tt}&=& \delta \epsilon \nonumber\\
\delta T^{ti}&=&(\epsilon + p)\delta u^{i}-\left(\frac{\alpha_{1}\varrho}{\epsilon + p} +2\alpha_{2}\right)\delta E^{i}+\delta^{ij}\left[ \frac{\alpha_{1}}{\epsilon + p}\partial_{j}\delta p + 2\alpha_{2}T \partial_{j}\left( \delta \frac{\mu}{T}\right) \right] -2\gamma_{B}\mu T^{2}\varepsilon^{tijk}\partial_{j}\delta u_{k}\nonumber\\
\delta T^{it}&=&(\epsilon + p)\delta u^{i} \nonumber\\  
\delta T^{ij}&=&\delta^{ij}\delta p -\eta \delta^{ik}\delta^{jm}\left(\partial_{k}\delta u_{m}+ \partial_{m}\delta u_{k}-\frac{2}{3}\delta_{km}\partial_{p}\delta u_{p}\right) -\zeta \delta^{ij}\partial_{k}\delta u_{k}=\delta T^{ji}\nonumber\\
\delta J^{t}&=& \delta \varrho +\frac{1}{2}\sigma_{B}\mathfrak{B}\delta u_{z}\nonumber\\
\delta J^{x}&=&\varrho \delta u_{x}+\left( \sigma_{E}+2\alpha_{3}\frac{\varrho}{\epsilon +p}\right)\delta E^{x}-\frac{2\alpha_{3}}{\epsilon + p}\partial_x \delta p - \sigma_{E}T \partial_{x}\left(\delta \frac{\mu}{T} \right)+\sigma_{V}\partial_{[y}\delta u_{z]}\nonumber\\
\delta J^{y}&=&\varrho \delta u_{y}+\left( \sigma_{E}+2\alpha_{3}\frac{\varrho}{\epsilon +p}\right)\delta E^{y}-\frac{2\alpha_{3}}{\epsilon + p}\partial_y \delta p - \sigma_{E}T \partial_{y}\left(\delta \frac{\mu}{T} \right)+\sigma_{V}\partial_{[z}\delta u_{x]}\nonumber\\
 \delta J^{z}&=&\varrho \delta u_{z}+\left( \sigma_{E}+2\alpha_{3}\frac{\varrho}{\epsilon +p}\right)\delta E^{z}-\frac{2\alpha_{3}}{\epsilon + p}\partial_z \delta p- \sigma_{E}T \partial_{z}\left(\delta \frac{\mu}{T} \right)+\sigma_{V}\partial_{[x}\delta u_{y]}+\delta \sigma_{B}\mathfrak{B}.\label{E25}
\end{eqnarray}
Note that the result $ \delta T^{ti} \neq \delta T^{it} $ essentially reflects the fact that we are working with the Lifshitz isometry group where the Lorentz boost invariance is explicitly broken.

Before we proceed further a few important remarks are in order. First of all, one should note that a Lifshitz fixed point is manifestly translation invariant. In other words, the Lifshitz symmetry algebra includes the generators of time translation ($ P_{t}=\partial_{t} $) as well as spatial translations ($ P_{i}=\partial_{i} $) that satisfy the following subalgebra with the dilatation generator ($ D $),
\begin{eqnarray}
[D,P_t]=z P_t,~~[D,P_i]=P_i.
\end{eqnarray} 
This eventually implies that in the vicinity of the Lifshitz like fixed points one might encounter with situations like infinite DC conductivity (due to the lack of momentum dissipation) that one generally encounters in the context of usual relativistic hydrodynamics due to the conservation of the linear momentum. The reason for this is that for a transnationally invariant system with finite charge density, the charge carriers could be accelerated upto infinite momentum in the presence of an external electric field \cite{Landsteiner:2014vua}. Such a situation could however be avoided once we include the effect of dissipation in the theory namely \cite{Landsteiner:2014vua}, 
\begin{eqnarray}
\partial_{\mu}\delta T^{\mu t}&=&\delta F^{tz}J_{z}+\frac{1}{\tau_{e}}\delta T^{\mu t}u_{\mu}\nonumber\\
\partial_{\mu}\delta T^{\mu i}&=&\varrho \delta F^{ti}+F^{i \mu}\delta J_{\mu}+\frac{1}{\tau_{m}}\delta T^{\mu i}u_{\mu}\nonumber\\
\partial_{\mu}\delta J^{\mu}&=& \mathfrak{c}\delta E^{\mu} B_{\mu}+\frac{1}{\tau_{c}}\delta J^{\mu}u_{\mu}\label{E26}
\end{eqnarray}
where, $ \tau_{e} $, $ \tau_{m} $ and $ \tau_{c} $ are respectively the energy relaxation time, momentum relaxation time and the charge relaxation time. In general for systems with spatial anisotropy the value of $ \tau_{m} $ should in principle be different along different spatial directions. However, for Lifshitz like fixed points we do not really need to worry about it as Lifshitz fixed points are manifestly rotationally invariant. Our next task would be to substitute (\ref{E25}) into (\ref{E26}) and restrict ourselves upto leading order in the fluctuations which finally yields, 
\begin{eqnarray}
\left(\partial_{t}+\frac{1}{\tau_{e}} \right)\delta \epsilon + \partial_{i}\left[(\epsilon + p)\delta u_{i} \right]-\sigma_{B} \mathfrak{B}\delta E_{z}=0\nonumber\\
\left(\partial_{t}+\frac{1}{\tau_{m}} \right)\left((\epsilon + p) \delta u_{x}-\Xi_{x}-4\gamma_{B}\mu T^{2}\partial_{[y}\delta u_{z]}\right)+ \partial_{x}\delta p - \eta \left(\partial_{k}^{2} \delta u_{x}+\frac{1}{3}\partial_{x}\partial_{k}\delta u_{k} \right)-\zeta \partial_{x}\partial_{k}\delta u_{k}\nonumber\\
=\varrho \delta F^{tx}+ \mathfrak{B}\delta J_{y}\nonumber\\
\left(\partial_{t}+\frac{1}{\tau_{m}} \right)\left((\epsilon + p) \delta u_{y}-\Xi_{y}-4\gamma_{B}\mu T^{2}\partial_{[z}\delta u_{x]}\right)+ \partial_{y}\delta p - \eta \left(\partial_{k}^{2} \delta u_{y}+\frac{1}{3}\partial_{y}\partial_{k}\delta u_{k} \right)-\zeta \partial_{y}\partial_{k}\delta u_{k}\nonumber\\
=\varrho \delta F^{ty}- \mathfrak{B}\delta J_{x}\nonumber\\
\left(\partial_{t}+\frac{1}{\tau_{m}} \right)\left((\epsilon + p) \delta u_{z}-\Xi_{z}-4\gamma_{B}\mu T^{2}\partial_{[x}\delta u_{y]}\right)+ \partial_{z}\delta p - \eta \left(\partial_{k}^{2} \delta u_{z}+\frac{1}{3}\partial_{z}\partial_{k}\delta u_{k} \right)-\zeta \partial_{z}\partial_{k}\delta u_{k}\nonumber\\
=\varrho \delta E_{z}\nonumber\\
\left(\partial_{t}+\frac{1}{\tau_{c}} \right)\left(\delta \varrho +\frac{1}{2}\sigma_{B}\mathfrak{B}\delta u_{z}\right)+ \partial_{i}\Upsilon_{i} +\partial_{z}\delta\sigma_{B}\mathfrak{B} =\mathfrak{c}\mathfrak{B} \delta E_{z}\nonumber\\
\label{E27}
\end{eqnarray}
where each of the individual coefficients could be formally expressed as ($ i = x,y,z $),
\begin{eqnarray}
\Xi_{i}&=&\left(\frac{\alpha_{1}\varrho}{\epsilon + p} +2\alpha_{2}\right)\delta E_{i}-\left( \frac{\alpha_{1}}{\epsilon + p}\partial_{i}\delta p + 2\alpha_{2}T \partial_{i}\left( \delta \frac{\mu}{T}\right) \right)\nonumber\\
&=&\left(\frac{\alpha_{1}\varrho}{\epsilon + p} +2\alpha_{2}\right)\delta E_{i}-\left(\frac{\alpha_{1}\varrho}{\epsilon + p} +2\alpha_{2}\right)\partial_i\delta \mu_{i}+..~..\nonumber\\
\Upsilon_{i}&=&\varrho \delta u_{i}+\left( \sigma_{E}+2\alpha_{3}\frac{\varrho}{\epsilon +p}\right)\delta E_{i}-\frac{2\alpha_{3}}{\epsilon + p}\partial_i \delta p - \sigma_{E}T \partial_{i}\left(\delta \frac{\mu}{T} \right)+\frac{\sigma_{V}}{2}\varepsilon^{ijk}\partial_{j}\delta u_{k}\nonumber\\
&=&\left( \sigma_{E}+2\alpha_{3}\frac{\varrho}{\epsilon +p}\right)\delta E_{i}-\left( \sigma_{E}+2\alpha_{3}\frac{\varrho}{\epsilon +p}\right)\partial_i\delta \mu_{i}+..~..
\end{eqnarray}
Before we proceed further, it is important to note that in the parity even sector the coefficients associated with $ \delta E_{i} $ as well as $ -\partial_i\delta \mu_{i} $ are indeed the same. The only difference that appears is in the parity odd sector, namely in the coefficients associated with $ \delta E_{z} $ which are vividly distinct in the last equation of (\ref{E27}). In other words, in the absence of anomaly, there would not have been any mismatch between the coefficients associated with $ \delta E_{i} $ and $ -\partial_i\delta \mu_{i} $.

Performing first the Fourier transform for the fluctuations and thereby taking the zero limit for the spatial momentum ($ \textbf{k} \rightarrow 0$) at the end we finally arrive at the following set of equations,
\begin{eqnarray}
\mathfrak{w}_{e}\delta \epsilon -i\sigma_{B}\mathfrak{B}\delta E_{z}&=&0\nonumber\\
\mathfrak{w}_{m}\left[ (\epsilon + p)\delta u_{x}-\left(\frac{\alpha_{1}\varrho}{\epsilon + p} +2\alpha_{2}\right)\delta E_{x}\right]-i\varrho \delta F^{tx}-i \mathfrak{B}\left[\varrho \delta u_{y}+\left( \sigma_{E}+2\alpha_{3}\frac{\varrho}{\epsilon +p}\right)\delta E_{y} \right]&=&0\nonumber\\
\mathfrak{w}_{m}\left[ (\epsilon + p)\delta u_{y}-\left(\frac{\alpha_{1}\varrho}{\epsilon + p} +2\alpha_{2}\right)\delta E_{y}\right]-i\varrho \delta F^{ty}+i \mathfrak{B}\left[\varrho \delta u_{x}+\left( \sigma_{E}+2\alpha_{3}\frac{\varrho}{\epsilon +p}\right)\delta E_{x} \right]&=&0\nonumber\\
\mathfrak{w}_{m}\left[ (\epsilon + p)\delta u_{z}-\left(\frac{\alpha_{1}\varrho}{\epsilon + p} +2\alpha_{2}\right)\delta E_{z}\right]-i\varrho \delta E_{z} &=&0\nonumber\\ 
\mathfrak{w}_{c}\left(\delta \varrho +\frac{1}{2}\sigma_{B}\mathfrak{B}\delta u_{z}\right) -i \mathfrak{c}\mathfrak{B}\delta E_{z}&=&0\nonumber\\ 
\end{eqnarray}  
where the individual frequencies could be formally expressed as,
\begin{eqnarray}
\mathfrak{w}_{e} = \mathfrak{w}+\frac{i}{\tau_{e}},~~\mathfrak{w}_{m} = \mathfrak{w}+\frac{i}{\tau_{m}},~~\mathfrak{w}_{c} = \mathfrak{w}+\frac{i}{\tau_{c}}.
\end{eqnarray}

Before we proceed further, it is important to note down the following variations namely,
\begin{eqnarray}
\delta \epsilon (\mu , T)& =& \mathfrak{g}_{1}\delta\mu +\mathfrak{g}_{2}\delta T \nonumber\\
\delta \varrho (\mu , T)& =& \mathfrak{h}_{1}\delta\mu +\mathfrak{h}_{2}\delta T.\label{E31}
\end{eqnarray}

By means of the above set of identities (\ref{E31}), our next task would be to solve the fluctuations namely $ \delta u_{z} $, $ \delta \mu $ and $ \delta T $ in terms the variations of the electric field strength which finally yields,
\begin{eqnarray}
\delta u_{z}&=&\frac{i \varrho \delta E_{z} }{\mathfrak{w}_{m}(\epsilon +p)}+\left( \frac{\alpha_{1}\varrho}{\epsilon + p}+2\alpha_{2}\right)\frac{ \delta E_{z} }{(\epsilon +p)}\nonumber\\
\delta \mu &=& \frac{\mathfrak{B}\delta E_{z}}{(\mathfrak{g}_{2}\mathfrak{h}_{1}-\mathfrak{g}_{1}\mathfrak{h}_{2})}\left(-\frac{ i \mathfrak{h}_{2}\sigma_{B}}{\mathfrak{w}_{e}}-\frac{i\sigma_{B}\mathfrak{g}_{2}\varrho}{2 \mathfrak{w}_{m}(\epsilon +p)} +\frac{i \mathfrak{c}\mathfrak{g}_{2}}{\mathfrak{w}_{c}}\right)-\left( \frac{\alpha_{1}\varrho}{\epsilon + p}+2\alpha_{2}\right)\frac{\mathfrak{B}\sigma_{B}\mathfrak{g}_{2}\delta E_{z}}{2(\epsilon +p)(\mathfrak{g}_{2}\mathfrak{h}_{1}-\mathfrak{g}_{1}\mathfrak{h}_{2})}\nonumber\\ 
\delta T &=& \frac{\mathfrak{B}\delta E_{z}}{(\mathfrak{g}_{2}\mathfrak{h}_{1}-\mathfrak{g}_{1}\mathfrak{h}_{2})}\left(\frac{ i \mathfrak{h}_{1}\sigma_{B}}{\mathfrak{w}_{e}}+\frac{i\sigma_{B}\mathfrak{g}_{1}\varrho}{2 \mathfrak{w}_{m}(\epsilon +p)} -\frac{i \mathfrak{c}\mathfrak{g}_{1}}{\mathfrak{w}_{c}}\right)+\left( \frac{\alpha_{1}\varrho}{\epsilon + p}+2\alpha_{2}\right)\frac{\mathfrak{B}\sigma_{B}\mathfrak{g}_{1}\delta E_{z}}{2(\epsilon +p)(\mathfrak{g}_{2}\mathfrak{h}_{1}-\mathfrak{g}_{1}\mathfrak{h}_{2})}.\nonumber\\
\label{E32}
\end{eqnarray}
Before we proceed further, the reader should be able to figure out the basic differences between the set of solutions (\ref{E32}) obtained at a Lifshitz fixed point to that with the earlier observations made in the context of usual relativistic hydrodynamics \cite{Landsteiner:2014vua}. From (\ref{E32}), it is in fact quite trivial to note that corresponding to each of the individual fluctuations we always have an extra contribution whose origin could be understood as the lack of boost invariance at a Lifshitz fixed point . Therefore the above set of solutions (\ref{E32}) might be regarded as a special class of solutions those are valid particularly in the context of Lifshitz hydrodynamics.

Finally, substituting (\ref{E32}) into the last equation of (\ref{E25}) and thereby considering the $ \textbf{k}\rightarrow 0 $ limit we note the following, 
\begin{eqnarray}
\delta J_{z}&=&\varrho \delta u_{z}+\left( \sigma_{E}+2 \alpha_{3}\frac{\varrho}{\epsilon + p}\right) \delta E_{z}+\mathfrak{B}\Phi^{(\mu)}\delta \mu +\mathfrak{B}\Phi^{(T)}\delta T\nonumber\\
&\equiv & \sigma_{DC}\delta E_{z}
\end{eqnarray}
where each of the individual coefficients could be formally expressed as,
\begin{eqnarray}
\Phi^{(\mu)}&=& \mathfrak{c}-\frac{\mathfrak{h}_{1}(\mathfrak{c}\mu^{2}-2\gamma_{B}T^{2})+2 \mathfrak{c}\mu \varrho}{2(\epsilon +p)}+\frac{\varrho (\mathfrak{g}_{1}+\varrho)(\mathfrak{c}\mu^{2}-2\gamma_{B}T^{2})}{2(\epsilon + p)^{2}}\nonumber\\
\Phi^{(T)}&=&-\frac{(\mathfrak{c}\mu^{2}-2\gamma_{B}T^{2})\mathfrak{h}_{2}-4\gamma_{B}T\varrho}{2(\epsilon + p)}+\frac{\varrho(\mathfrak{c}\mu^{2}-2\gamma_{B}T^{2})(\mathfrak{g}_{2}+s)}{2(\epsilon +p)^{2}}.
\end{eqnarray}
Here $ \sigma_{DC} $ is the anomalous DC conductivity at a Lifshitz fixed point that could be formally expressed as,
\begin{eqnarray}
\sigma_{DC}=\sigma_{E}+\Theta +\Theta_{\mathfrak{L}}\label{E35}
\end{eqnarray}
where, $ \Theta $ is the usual contribution to the anomalous DC conductivity corresponding to $ z=1 $ fixed point \cite{Landsteiner:2014vua}. On the other hand, $ \Theta_{\mathfrak{L}} $ is the non trivial contribution to the conductivity that appears solely due to the Lifshitz scaling symmetry. In this sense (\ref{E35}) is the generalization of the earlier observations \cite{Landsteiner:2014vua} corresponding to Lifshitz like fixed points.  The details of these coefficients could be formally expressed as,
\begin{eqnarray}
\Theta =\frac{i \varrho^{2}}{\mathfrak{w}_{m}(\epsilon +p)}+\frac{\mathfrak{B}^{2}\Phi^{(\mu)}}{(\mathfrak{g}_{2}\mathfrak{h}_{1}-\mathfrak{g}_{1}\mathfrak{h}_{2})}\left(-\frac{ i \mathfrak{h}_{2}\sigma_{B}}{\mathfrak{w}_{e}}-\frac{i\sigma_{B}\mathfrak{g}_{2}\varrho}{2 \mathfrak{w}_{m}(\epsilon +p)} +\frac{i \mathfrak{c}\mathfrak{g}_{2}}{\mathfrak{w}_{c}}\right)\nonumber\\
+\frac{\mathfrak{B}^{2}\Phi^{(T)}}{(\mathfrak{g}_{2}\mathfrak{h}_{1}-\mathfrak{g}_{1}\mathfrak{h}_{2})}\left(\frac{ i \mathfrak{h}_{1}\sigma_{B}}{\mathfrak{w}_{e}}+\frac{i\sigma_{B}\mathfrak{g}_{1}\varrho}{2 \mathfrak{w}_{m}(\epsilon +p)} -\frac{i \mathfrak{c}\mathfrak{g}_{1}}{\mathfrak{w}_{c}}\right)
\label{E36}
\end{eqnarray}
\begin{eqnarray}
\Theta_{\mathfrak{L}}=\left( \frac{\alpha_{1}\varrho}{\epsilon + p}+2\alpha_{2}\right)\left(\frac{\varrho}{\epsilon + p} -\frac{\mathfrak{B}^{2}\Phi^{(\mu)}\sigma_{B}\mathfrak{g}_{2}}{2(\epsilon + p)(\mathfrak{g}_{2}\mathfrak{h}_{1}-\mathfrak{g}_{1}\mathfrak{h}_{2})}+\frac{\mathfrak{B}^{2}\Phi^{(T)}\sigma_{B}\mathfrak{g}_{1}}{2(\epsilon + p)(\mathfrak{g}_{2}\mathfrak{h}_{1}-\mathfrak{g}_{1}\mathfrak{h}_{2})}\right)+\frac{2 \alpha_{3}\varrho}{\epsilon + p}.\nonumber\\
\label{E37} 
\end{eqnarray}

Eq.(\ref{E35}) along with Eqs. (\ref{E36}) and (\ref{E37}) provides a complete description for the anomalous conductivity corresponding to Lifshitz like fixed points. In the following we systematically enumerate a number of interesting observations one by one.\\ \\
$\bullet$ In the limit $ \mathfrak{B},\mathfrak{c},\gamma_{B}\rightarrow 0 $, the conductivity along the longitudinal direction turns out to be,
\begin{eqnarray}
\sigma_{DC}=\sigma_{E}+\frac{i \varrho^{2}}{\mathfrak{w}_{m}(\epsilon + p)}+\left( \frac{\alpha_{1}\varrho}{\epsilon + p}+2\alpha_{2}+2\alpha_{3}\right)\frac{\varrho}{\epsilon + p}.\label{E38}
\end{eqnarray}
As usual the first two terms on the R.H.S. of (\ref{E38}) corresponds to the usual contribution to the anomalous conductivity in the absence of the background magnetic field ($ \mathfrak{B} $)\cite{Landsteiner:2014vua}. The last term on the R.H.S. of (\ref{E38}) is the contribution that solely arises because of the Lifshitz scaling symmetry. The source for the first two terms are hidden in the constitutive relation for the stress tensor ($ T^{\mu \nu} $) while the third term has its origin in the acceleration piece appearing in the constitutive relation of the $ U(1) $ current ($ J^{\mu} $).\\ \\
$ \bullet $ One could make various important observations by looking at the expression for $ \Theta_{\mathfrak{L}}$ quite carefully\footnote{In this paper we are not interested in the piece $ \Theta $ as it has been already discussed extensively in \cite{Landsteiner:2014vua}.}. The first observation that one should be able to make is that clearly there seems to be a precise contribution of the anomaly sector on the charge current which is always coupled to the background magnetic field ($ \mathfrak{B} $). The most amazing fact about the Lifshitz contribution to the longitudinal DC conductivity ($ \sigma_{DC} $) is that unlike the $ z=1 $ case, even in the absence of the anomaly ($ \mathfrak{c}=0 $), we have a precise contribution to $ \sigma_{DC} $ sourced by the external magnetic field ($ \mathfrak{B} $) whose strength is determined by the coefficient $ \gamma_{B} $ which is the parity odd transport associated with the axial vector ($ \omega^{\mu} $) in the constitutive relation for the stress tensor (\ref{E19}). One could separate out this parity odd contribution to the magnetoconductivity as,
\begin{eqnarray}
\Theta_{\mathfrak{L}}=\left( \frac{\alpha_{1}\varrho}{\epsilon + p}+2\alpha_{2}\right)\frac{\mathfrak{B}^{2}\varrho T^{4}\gamma_{B}^{2}}{2(\epsilon +p)^{3}(\mathfrak{g}_{2}\mathfrak{h}_{1}-\mathfrak{g}_{1}\mathfrak{h}_{2})}\left(\mathfrak{g}_{2}\mathfrak{h}_{1}-\mathfrak{g}_{1}\mathfrak{h}_{2} -\frac{\varrho(\mathfrak{g}_{2}\varrho -\mathfrak{g}_{1}s)}{\epsilon +p}-\frac{2\mathfrak{g}_{1}\varrho}{T}\right). 
\end{eqnarray}
At this stage it is noteworthy to mention that for the $ z=1 $ case, the effect of the external magnetic field ($ \mathfrak{B} $) to the longitudinal conductivity ($ \sigma_{DC} $) is always coupled with the anomaly ($ \mathfrak{c} $) itself \cite{Landsteiner:2014vua}. On the other hand, as we have just seen, in the Lifshitz scenario the anomaly ($ \mathfrak{c} $) is not the only candidate that contributes to the conductivity via external magnetic field ($ \mathfrak{B} $). All these observations eventually suggest that for Lifshitz like systems, even in the absence of the anomaly, one could in principle have a finite change in the chemical potential ($ \delta \mu $) or in the temperature ($ \delta T $) associated with the external magnetic field.\\ \\
$ \bullet $ Like in the sector $ \Theta $, one could also divide various contributions appearing in the Lifshitz sector $ \Theta_{\mathfrak{L}}$ mostly into three pieces. The first term is the usual contribution that appears due to the acceleration ($ \delta u_{z} $) of the charged particles in the presence of an external electric filed ($ \delta E_{z} $). The second and the third pieces are precisely the contributions that appear due to the change in the chemical potential ($ \delta \mu $) as well as the temperature ($ \delta T $) of the system while switching on the external electric field ($ \delta E_{z} $). These last two effects are precisely the Lifshitz sector of the anomalous contributions to the longitudinal current ($ \delta J_{z} $).\\\\
$ \bullet $ The significant difference that appears in the Lifshitz sector ($ \Theta_{\mathfrak{L}} $) of the anomalous conductivity is that unlike the previously explored \cite{Landsteiner:2014vua} sector $ \Theta $, the Lifshitz sector ($ \Theta_{\mathfrak{L}} $) does not explicitly depend on the frequency ($ \mathfrak{w} $) and it does not contain any pole in $ \mathfrak{w} $. Therefore this piece is always finite at all frequencies. In other words, from Lifshitz sector one can always have a finite contribution to the longitudinal conductivity irrespective of any frequency. Moreover, it also does not contain any information about the relaxation times present in the system and therefore is completely blind to any effects of dissipation present in the system.\\ \\
$ \bullet $ Finally, we would like to mention about the last piece of information available above in (\ref{E37}) namely, $ \frac{2 \alpha_{3}\varrho}{\epsilon + p} $ which could be thought of as an effect arising due to the acceleration of the charged particles in the system. We identify this effect as the consequence of the acceleration term present in the constitutive relation of the charge current (\ref{E20}).

\section{A note on holographic derivation}
The purpose of this section is to make an outline of the possible set up that will be required in order to carry out an explicit holographic computation for the physical entity $ \Theta_{\mathfrak{L}} $ in the \textit{probe} limit. With these calculations in hand, we shall atleast have some prescription in order to evaluate the actual value of the anomalous contribution of the Lifshitz sector to the longitudinal DC conductivity. The computation that we carry out in this section eventually assumes the following facts:\\
$ \bullet $ The gauge fields ($\mathsf{A}_{\mu} $) are considered to be in the probe limit, i.e, they do not back react on the background geometry.\\
$ \bullet $ The charge density ($ \varrho $) of the system is considered to be extremely small compared to the neutral d.o.f of the system. On the other hand, the temperature ($ T $) of our system turns out to be extremely high so that the following limits hold true namely, $ \frac{\partial \varrho}{\partial T}\ll 1 $, $ |\mathfrak{c}\mathfrak{B}|\ll T^{2} $ and $ \mu/T \ll 1 $.
With these assumptions in mind, the approximate expression for $ \Theta_{\mathfrak{L} }$ turns out to be,
\begin{eqnarray}
\Theta_{\mathfrak{L}}\approx \frac{\alpha_{2}\mathfrak{B}^{2}\sigma_{B}\gamma_{B}}{s^{2}}\left(\frac{\mathfrak{g}_{1}\mathfrak{h}_{2}}{\mathfrak{g}_{2}\mathfrak{h}_{1}}-1 \right). 
\label{E42}
\end{eqnarray}
Our job would be to evaluate the above entity (\ref{E42}) using the techniques of Gauge/gravity duality.
\subsection{The gravity set up}
The gravity set up for our present calculation in the bulk essentially consists of asymptotically (uncharged) Lifshitz black hole solutions in $ (4+1) $ dimensions. As observed in \cite{Taylor:2008tg}, the effective action for Lifshitz like black brane solutions could be formally expressed as,
\begin{eqnarray}
S=\frac{1}{16 \pi G_{5}}\int d^{5}x \sqrt{-g}\left( R -2\Lambda -\frac{1}{2}\partial_{\mu}\phi \partial^{\mu}\phi -\frac{1}{4}e^{\lambda \phi}\mathcal{F}_{\mu \nu}\mathcal{F}^{\mu \nu}\right)
\label{E43} 
\end{eqnarray}
where, $ \phi $ is the mass less scalar field, $ \mathcal{F} $ is the filed strength tensor corresponding to an abelian one form $ (\mathcal{A}_{\mu})  $ and $ \Lambda $ is the negative cosmological constant.

The resulting Lifshitz black brane solution that naturally emerges as a solution of (\ref{E43}) could be formally expressed as \cite{Pang:2009wa},
\begin{eqnarray}
ds^{2}&=&L^{2}\left( -r^{2z}f(r)dt^{2}+\frac{dr^{2}}{r^{2}f(r)}+r^{2}d \textbf{x}^{2}\right) \nonumber\\
f(r)&=&1-\frac{r_{0}^{z+3}}{r^{z+3}},~~\phi (r)\sim \log r,~~\mathcal{F}_{rt}\sim L r^{z+2}\nonumber\\
\Lambda &=& -\frac{(z+3)(z+2)}{2 L^{2}}.
\label{E44}
\end{eqnarray}
Before we proceed further, it is customary to note down the following important facts. First of all, here $ r_{0} $ denotes the location of the horizon of the black brane. Whereas, on the other hand, the boundary of the space time is located at $ r \rightarrow \infty $. Finally, the Hawking temperature ($ T $), entropy density ($ s $) as well as the energy density ($ \epsilon $) for such space time configuration turn out to be,
\begin{eqnarray}
T = \frac{(z+3)r_{0}^{z}}{4 \pi},~~s=\frac{L^{3}r_{0}^{3}}{4 G_{5}},~~\epsilon = \frac{3L^{3}r_{0}^{z+3}}{16 \pi G_{5}}.
\label{E45}
\end{eqnarray}
\subsection{Calculation of $ \Theta_{\mathfrak{L}} $}
To start with, we define a new variable,
\begin{eqnarray}
u = \frac{r_{0}}{r}
\end{eqnarray}
in terms of which the solution (\ref{E44}) turns out to be,
\begin{eqnarray}
ds^{2}&=&L^{2}\left( -\left( \frac{r_{0}}{u}\right) ^{2z}f(u)dt^{2}+\frac{du^{2}}{u^{2}f(u)}+\left( \frac{r_{0}}{u}\right) ^{2}d \textbf{x}^{2}\right) \nonumber\\
f(u)&=&1-u^{z+3}.
\label{E47}
\end{eqnarray}
Note that in this coordinate system, the horizon of the Lifshitz black brane is located at $ u=1 $, whereas on the other hand, the boundary is located at $ u=0 $.

In order to describe the boundary (Lifshitz) hydrodynamics of a charged anomalous fluid in ($ 3+1 $) dimensions in the small charge density ($ \varrho \ll TL $) limit we consider the Maxwell Chern-Simons (CS) action,
\begin{eqnarray}
\mathsf{S}=\int d^{5}x\sqrt{-g}\left( -\frac{1}{4}\mathsf{F}^{2}+\frac{\kappa}{3}\frac{\varepsilon^{\mu \nu \rho \sigma \lambda}}{\sqrt{-g}}\mathsf{A}_{\mu}\mathsf{F}_{\nu\rho}\mathsf{F}_{\sigma\lambda}\right) 
\end{eqnarray}
as a probe over the neutral background (\ref{E47}). Furthermore, from now on we set $ L=16\pi G_{5}=1 $ for the rest of our analysis. The resulting equation of motion turns out to be, 
\begin{eqnarray}
\nabla_{\sigma}\mathsf{F}^{\sigma \lambda}+\frac{\kappa}{\sqrt{-g}}\varepsilon^{\lambda \alpha\beta\gamma\delta}\mathsf{F}_{\alpha\beta}\mathsf{F}_{\gamma\delta}=0.
\label{E49}
\end{eqnarray}

In order to solve (\ref{E49}), we choose the following ansatz for the gauge field namely,
\begin{eqnarray}
\mathsf{A}_{\mu}=(\varphi (u),0,0,\mathfrak{B}x,\mathsf{A}_{z}(u)).
\label{E50}
\end{eqnarray}

Substituting the above ansatz (\ref{E50}) into (\ref{E49}), we arrive at the following set of equations\footnote{We have set $ \varepsilon^{txyzu}=1 $.},
\begin{eqnarray}
\varphi''(u)+\frac{(z-2)}{u}\varphi'(u)+\frac{8\kappa \mathfrak{B}}{r_{0}^{3-z}u^{z-2}}\mathsf{A}'_{z}&=&0\nonumber\\
\mathsf{A}_{z}'' +\left(\frac{f'(u)}{f(u)}-\frac{z}{u} \right)\mathsf{A}_{z}'+\frac{8\kappa\mathfrak{B}u^{z}}{r_{0}^{z+3}f(u)}\varphi'(u)&=&0.
\label{E51} 
\end{eqnarray}

The above set of equations (\ref{E51}) could be formally re expressed as,
\begin{eqnarray}
\left( \frac{8 \kappa\mathfrak{B}}{r_{0}^{3-z}}\mathsf{A}_{z}+u^{z-2}\varphi'(u)\right) '&=&0\nonumber\\
\frac{f(u)}{u^{z}}\mathsf{A}_{z}' +\frac{8\kappa\mathfrak{B}}{r_{0}^{z+3}}\varphi(u)&=&0
\label{E52}
\end{eqnarray}
subjected to the fact that $ \varphi(1)=f(1)=0 $. Combining the above two equations in (\ref{E52}) we finally obtain,
 \begin{eqnarray}
 \varphi''(u)+\frac{(z-2)}{u}\varphi'(u)-\frac{64 \kappa^{2}\mathfrak{B}^{2}u^{2}}{r_{0}^{6}(1-u^{z+3})}\varphi(u)=0.
 \label{E53}
 \end{eqnarray}
The above differential equation (\ref{E53}) in general is quite difficult to solve for some generic values of the dynamic critical exponent ($ z $). Therefore in order to solve (\ref{E53}) analytically we choose, $ z=2 $ Lifshitz fixed point for the present case of study. With this choice, the corresponding near boundary solution ($ u\rightarrow 0 $) turns out to be,
\begin{eqnarray}
\varphi (u)= \frac{\sqrt{\pi }}{\sqrt[4]{2} \Gamma \left(\frac{3}{4}\right)}-\frac{4 u \left(\sqrt[4]{2} \sqrt{\pi } \sqrt{\mathfrak{B}} \sqrt{\kappa }\right)}{r_{0}^{3/2}\Gamma \left(\frac{1}{4}\right)} +\mathcal{O}(u^{2}).
\end{eqnarray}
 The chemical potential ($ \mu $) for the boundary field theory turns out to be,
\begin{eqnarray}
\mu = \varphi (0)=\frac{\sqrt{\pi }}{\sqrt[4]{2} \Gamma \left(\frac{3}{4}\right)}.
\label{E55}
\end{eqnarray}
On the other hand, the charge density ($ \varrho $) for the boundary field theory turns out to be,
\begin{eqnarray}
\varrho =\frac{4  \left(\sqrt[4]{2} \sqrt{\pi } \sqrt{\mathfrak{B}} \sqrt{\kappa }\right)}{r_{0}^{3/2}\Gamma \left(\frac{1}{4}\right)}.
\label{E56}
\end{eqnarray}

Using (\ref{E45}), (\ref{E55}) and (\ref{E56}) we finally obtain,
\begin{eqnarray}
\Theta_{\mathfrak{L}}\approx \frac{2600 \alpha_{2}\mathfrak{B}^{2}\gamma_{B}}{2^{1/4}(4\pi)^{3}T^{3}}\left( \frac{5^{9/4}\sqrt{\mathfrak{B}\kappa}\gamma_{B}}{\pi^{7/4}\Gamma (1/4)}-\frac{\mathfrak{c}\sqrt{\pi}}{\Gamma (3/4)}\right).
\label{E57}
\end{eqnarray}
Note that even if $ \mathfrak{c}=0 $, we still have a precise contribution to the longitudinal DC conductivity ($ \sigma_{DC} $) via this additional parity odd transport $ \gamma_{B} $. 

Following the original approach developed in \cite{Landsteiner:2012kd}, our final goal is to provide the Kubo formulae corresponding to two of the transport coefficients namely, $ \alpha_{2} $ and $ \gamma_{B} $. In order to do that, we consider the simplest situation namely we go to the rest frame of the fluid such that $ u^{\mu}=(1,0,0,0) $. Moreover, we turn on the vector potential only along the $ y $ direction and at the same time the metric fluctuation with only non vanishing component $ h_{ty} $ such that all of these fluctuations depend only along the $ z $ spatial direction. Considering all these facts we note that upto leading order in the fluctuations,
\begin{eqnarray}
T^{tx}&=&2\gamma_{B}\mu T^{2}\partial_{z}h_{ty}\nonumber\\
T^{ty}&=&-p h_{ty}-\alpha_{1}\partial_{t}h_{ty}+2\alpha_{2}\partial_{t}\mathsf{A_{y}}.
\label{E59}
\end{eqnarray} 
Finally, putting the above relations (\ref{E59}) into the momentum space and differentiating with respect to the sources we note down the Kubo formulae in the Landau frame as,
\begin{eqnarray}
\gamma_{B}&=&\lim_{\mathsf{k}\rightarrow 0}\frac{1}{2i\mathsf{k} \mu T^{2}}\langle T^{tx}T^{ty}\rangle \nonumber\\
\alpha_{2}&=&-\lim_{\mathfrak{w}\rightarrow 0}\frac{1}{2i\mathfrak{w}}\langle T^{ty}J^{y}\rangle .
\label{E61}
\end{eqnarray}
Note that in the Kubo formula of $ \gamma_{B} $ one needs to take the zero momentum ($ \mathsf{k}\rightarrow 0 $) limit followed by a zero frequency ($ \mathfrak{w}\rightarrow 0 $) limit which is quite reminiscent to that of the Kubo formula for anomalous transports \cite{Landsteiner:2012kd}. This similarity stems from the fact that both of these transports are associated with the parity odd contributions to the constitutive relations (\ref{E19}) and (\ref{E20}). On the other hand, $ \alpha_{2} $ is the usual dissipative transport which is also reflected in its Kubo formula (\ref{E61}).

We conclude our analysis with the following comments. First of all, it is interesting to note from (\ref{E57}) is that $ \Theta_{\mathfrak{L}} $ is finally determined in terms of only two of the Lifshitz transports namely, $ \alpha_{2} $ and $ \gamma_{B} $. The other two transports ($ \alpha_{1} $ and $ \alpha_{3} $) are not quite relevant in the low charge density limit. Secondly, and most importantly, we note that the leading behavior of $ \Theta_{\mathfrak{L}} $ goes with temperature as $ \sim T^{-5} $, which therefore suggests that the anomalous contribution to the charge current is highly suppressed in the high temperature limit. Contrary to the previous observations corresponding to the $ z=1 $ case (where the anomalous contribution to the conductivity scales as $ \sim T^{-2} $\cite{Landsteiner:2014vua}), this observation is in fact quite non trivial in the sense that for the Lifshitz sector the suppression of the conductivity with respect to the temperature is much more faster than its cousin in the relativistic sector.  

\section{Summary and final remarks}
In this paper, based on the framework of linear response theory, we perform an analytic computation for the longitudinal DC conductivity associated with Lifshitz like fixed points in the presence of chiral anomalies in ($ 3+1 $) dimensions. The key findings of our analysis could be summarized as follows:\\
$ \bullet $ Apart from having the usual contributions coming from the chiral anomaly ($ \mathfrak{c} $), in our analysis we discover an additional (parity odd) contribution to the magnetoconductivity whose origin could be traced back into the lack of Lorentz boost invariance at a Lifshitz fixed point.\\
$ \bullet $ The Lifshitz sector ($ \Theta_{\mathfrak{L}} $) of the DC conductivity does not contain any information regarding the various relaxation times present in the system. It is also independent of the frequency ($ \mathfrak{w} $) and therefore finite as $ \mathfrak{w}\rightarrow 0 $.\\
$ \bullet $ Finally, in our analysis we device the appropriate holographic set up in order to compute $ \Theta_{\mathfrak{L}} $ at strong coupling and low charge density limit. From our analysis we note that with the increase in temperature, $ \Theta_{\mathfrak{L}} $ decreases more rapidly compared to its relativistic ($ z=1 $) cousins.
\\ \\
{\bf {Acknowledgements :}}
 The author would like to acknowledge the financial support from UGC (Project No UGC/PHY/2014236).\\


\end{document}